\documentclass{aa} 
\usepackage{natbib,amssymb}
\bibpunct{(}{)}{;}{a}{}{,} 
\usepackage{graphicx}

\begin{document}

\title{Crystallinity versus mass-loss rate in Asymptotic Giant Branch
  stars}

\author{F.~Kemper\inst{1} \and L.B.F.M.~Waters\inst{1,2} \and
  A.~de~Koter\inst{1} \and A.G.G.M.~Tielens\inst{3,4}}

\institute{Astronomical Institute 'Anton Pannekoek', University of
  Amsterdam, Kruislaan 403, 1098 SJ Amsterdam, The Netherlands \and
  Instituut voor Sterrenkunde, Katholieke Universiteit Leuven,
  Celestijnenlaan 200B, 3001 Heverlee, Belgium \and SRON Laboratory
  for Space Research, P.O.~Box 800, 9700 AV Groningen, The Netherlands
  \and Kapteijn Institute, University of Groningen, P.O.~Box 800, 9700
  AV Groningen, The Netherlands}

\offprints{F.~Kemper (ciska@astro.uva.nl)}

\date{Received / Accepted}

\maketitle

\begin{abstract}
Infrared Space Observatory (ISO) observations have shown that O-rich
  Asymptotic Giant Branch (AGB) stars exhibit crystalline silicate
  features in their spectra only if their mass-loss rate is higher
  than a certain threshold value. Usually, this is interpreted as
  evidence that crystalline silicates are not present in the dust
  shells of low mass-loss rate objects. In this study, radiative
  transfer calculations have been performed to search for an
  alternative explanation to the lack of crystalline silicate features
  in the spectrum of low mass-loss rate AGB stars. It is shown that
  due to a temperature difference between amorphous and crystalline
  silicates it is possible to include up to 40\% of crystalline
  silicate material in the circumstellar dust shell, without the spectra
  showing the characteristic spectral features. Since this implies
  that low mass-loss rate AGB stars might also form crystalline
  silicates and deposit them into the Interstellar Medium (ISM), the
  described observational selection effect may put the process of dust
  formation around AGB stars and the composition of the predominantly
  amorphous dust in the
  Interstellar Medium in a different light.
  Our model calculations result in a diagnostic tool to determine
  the crystallinity of an AGB star with a known mass-loss rate.
  \keywords{stars: AGB and post-AGB -- circumstellar matter -- dust,
    extinction -- infrared: stars -- radiative transfer}
\end{abstract}

\section{Introduction}
\label{sec:intro}

Oxygen-rich 
Asymptotic Giant Branch (AGB) stars are M-type stars of intermediate mass ($1
M_{\odot} < M < 8 M_{\odot}$) in the late stages of stellar evolution,
which exhibit He- and H-shell burning 
(for a review on AGB stars, see \citet{H_96_review}, and references herein).
These stars undergo an
increasing mass loss while they evolve along the AGB; the stars with low
mass-loss rates ($\sim$ 10$^{-8}$ - $\sim$ 10$^{-6}$ $M_{\odot}$
yr$^{-1}$) are referred to as Miras, whereas the stars with high mass-loss 
rates ($\sim$ 10$^{-5}$ - $\sim$ 10$^{-3}$ $M_{\odot}$ yr$^{-1}$) are the
so-called OH/IR stars. In the cooling outflow dust formation occurs,
which adds an infrared excess to the Spectral Energy Distribution
(SED). In case of Miras, the dust shell is optically thin, and the
underlying stellar spectrum is still visible, but 
the optically thick dust shell of OH/IR stars 
completely obscures the stellar
spectrum, making these objects only detectable at infrared wavelengths
\citep{B_87_dustshells,VH_88_IRAScolors}.

Important reasons for studying the nature of dust in AGB stars can be
found in the fact that these stars are the main contributors of dust
into the interstellar medium. The properties of the dust in the
ISM can thus be directly compared to the dust
properties in shells around evolved stars. This will provide
information on the processing of the dust in the ISM and will improve
our insight into the life cycle of dust. Moreover, studying the
condensation of dust in the outflows of AGB stars will provide useful
diagnostics for the determination of the physical conditions in these
outflows, thus increasing our understanding of this important stage of
stellar evolution.

Most AGB stars exhibit an oxygen-rich chemistry, i.e.~a number ratio
C/O $<$ 1, in their
dust shells, which gives rise to the condensation of oxygen-rich dust
species. Based on early infrared observations, the presence of
silicate species has been proposed by several authors
\citep[e.g.][]{G_69_composition,JM_76_dust}. Pre-ISO studies almost
exclusively pointed toward the presence of amorphous Fe-bearing 
silicates; such dust shells are for example studied by
\citet{BW_83_oh26,PP_85_midIR} and \citet{B_87_dustshells}. 
Infrared spectroscopy
performed with the Infrared Space Observatory \citep{KSA_96_ISO}
revealed the presence of 
crystalline silicates next to the amorphous silicate dust in the circumstellar
dust shells of some evolved stars \citep{WMJ_96_mineralogy}. In addition,
the observed wavelengths of the emission bands of the crystalline
silicates in evolved stars suggest that these grains contain little or
no Fe, but are very Mg-rich \citep{MWT_99_afgl4106}.  The data are
consistent with no Fe in the grains.  There seems to be a correlation
between the peak of the SED, i.e.~the optical depth toward the central
star, and the presence of spectral features of crystalline olivine
(forsterite, Mg$_2$SiO$_4$) and crystalline pyroxene (enstatite,
MgSiO$_3$) \citep{WMJ_96_mineralogy,CJJ_98_ohir,SKB_99_ohir}, in the
sense that this material is not seen in stars that are not obscured.
This is usually interpreted as evidence that crystalline silicates are
not present in the dust shells around low mass-loss rate AGB stars,
which have an optically thin dust shell. Apparently, a
threshold value for the mass-loss rate $\dot{M}$ is required to enable
the condensation of crystalline silicate in the stellar outflows. This
critical density would thus put constraints on the dust condensation
sequence. Various explanations have been offered for this threshold density
\citep{TWM_98_mineralogy,GS_99_condensation,SK_99_xsilagb}. Yet, its
existence is not well established.

In this study, we will evaluate the existence of the
threshold value for the mass-loss rate below which crystalline silicates are
not detected.  Does the lack of crystalline silicate features in the
SED of Miras really imply that the crystalline silicates are
not present, or is some other effect at work suppressing the
characteristic narrow features? We will study this issue by
calculating a grid of detailed model spectra of circumstellar dust
shells for a wide range of mass loss and degree of crystallinity.

In Sect.~\ref{sec:model} the model for the circumstellar dust shell
will be described, including the applied optical constant for the
silicates. In Sect.~\ref{sec:results} the results will be discussed
for varying mass-loss rate and degree of crystallinity. 
Sect.~\ref{sec:discussion} contains the physical explanation of the obtained
results and highlights possible implications for the evolution of
dust. A summary of this study is given in Sect.~\ref{sec:summ}.

\section{Modelling the circumstellar dust shell}
\label{sec:model}

The dust radiative transfer programme {\sc modust} (de Koter et al.,
\emph{in prep.})  has been used to calculate detailed spectra of 
circumstellar dust shells. The code solves the transfer equations 
      subject to the constraint of radiative equilibrium,
      fixing the temperature distribution of the dust grains. The solution
      technique focusses on the combined moment form of the transfer equation, 
      such that the scattering term in the source function can be solved for 
      explicitely, updating the Eddington factors throughout the lambda
      iteration by means of a ray-by-ray formal solution.

 In the mode used in this paper, the dust is assumed to be distributed
      in a spherical shell with a radial density profile $\rho \propto r^{-2}$,
      i.e.~that of a time-independent stellar wind outflow at constant 
      velocity
      $$
         \dot{M} = 4\pi r^{2} \rho(r) v_{\rm exp}
      $$
      where $\dot{M}$ is the mass-loss rate and $v_{\rm exp}$ the
      outflow velocity.      
      The program allows one to specify an arbitrary number of dust species, 
      each with its own shape and size distribution 
      \citep[see also][]{BKA_00_dust}. Here we adopted spherical grains, 
      for which we use Mie 
      calculations to determine the absorption and scattering coefficients. 
      The adopted chemical composition and laboratory measurements of
      optical constants -- necessary to calculate the extinction properties --
      will be discussed in Sect.~\ref{sec:optconst}. 
      For the size distribution we used a 
      powerlaw $N(a) \propto a^{-3.5}$, consistent with the grain-size 
      distribution of the ISM \citep{MRN_77_grainsize}. 
Following studies on the sizes of interstellar
grains and of stardust recovered from meteorites 
\citep{AZ_93_stardust,KMH_94_sizedistr} we have adopted a grain-size range of
0.01 -- 1 $\mu$m.
The dust shell consists of separate grain populations, i.e.~each dust
grain contains one type of silicate, implying that the separate dust
species can have different temperature profiles. Finally, all dust
species have the same spatial distribution.

The models in this study cover a wide range in mass-loss rates and
degree of crystallinity. The mass-loss rate is varied from $5 \cdot
10^{-8}$ to $10^{-4}$ $M_{\odot}$ yr$^{-1}$. The crystallinity $x$,
defined as
$$
x = \frac{\textrm{total mass crystalline silicates}}{\textrm{total
    mass silicates}}
$$
is varied from 0\% to 50\%. The dust shell
 is illuminated
at the inner radius by a typical M9 giant, of which the
spectrum is taken from \citet{FPT_94_Mstars}. The
size of the inner radius $R_{\mathrm{in}}$ of the dust shell is
determined by the condensation temperature of the dust,
$T_{\mathrm{cond}} \approx 1000$ K \citep{GS_99_condensation}.  The
outer radius is set to $R_{\mathrm{out}} = 200 R_{\mathrm{in}}$, which
is sufficiently far out to account for all the near infrared (NIR) and 
mid infrared (MIR) flux.  The
outflow velocity is assumed to be at a constant value of
$v_{\mathrm{exp}} = 20$ km s$^{-1}$. The dust/gas mass ratio in the
outflow is taken to be $f = 0.01$.  The last two numbers are typical for
Miras and OH/IR stars. The presented results are expressed in terms
of mass-loss rates for the above value of the outflow velocity. However
the critical parameter determining the spectrum is obviously 
the dust density $\rho_{\mathrm{dust}} (r)$.
Therefore the results are invariant for the quantity 
$$
Q = \frac{\dot{M}}{v_{\mathrm{exp}}(1+f^{-1})} \simeq 
\frac{f \dot{M}}{v_{\mathrm{exp}}} \qquad \textrm{for} \quad f \ll 1
$$
which follows from the mass continuity equation.

\subsection{Dust optical constants}
\label{sec:optconst}

Previous studies of the SED of dust shells surrounding AGB stars
\citep{B_87_dustshells,JT_92_SED,LL_93_SED,LL_96_SED} used 
so-called \emph{astronomical} or \emph{dirty} silicates, of which the
optical constants were observationally determined
\citep{JM_76_dust,DL_84_optprop}.   The chemical
composition of astronomical silicate is not known, although likely the
extra absorption is due to the presence of iron in the form of Fe$^{2+}$ 
in the mineral \citep{ST_89_theory}.

In the last few years, accurate laboratory measurements of the optical
constants of amorphous silicates \citep{DBH_95_glasses} and
crystalline silicates
\citep{KST_93_olivpyro,KS_98_IRspectra,JMD_98_crystalline} have become
available. Together with the high spectral resolution of the ISO 
Short Wavelength Spectrometer (SWS) \citep{GHB_96_SWS} 
and Long Wavelength Spectrometer (LWS) \citep{CAA_96_LWS}  
this opens the possibility to determine
the dust composition in great detail, at least in principle. 
However, model calculations have
shown that the SED can not be fitted using only the optical
constants of amorphous and crystalline silicates. An additional source
of opacity in the NIR appears to be missing (Kemper et al.,
\emph{in prep}). This is a well-known problem which dates back to
\citet{JM_76_dust,B_87_dustshells} and \citet{ST_89_theory}. 
Although it is a fundamental issue, for the present study it
suffices to use the optical constants
of the amorphous silicates calculated from the observed spectra of AGB
stars \citep{S_99_optprop}.  The thus obtained optical constants of
the AGB silicate cannot provide information on the exact
composition of the amorphous component of the dust shell, however for
our studies of the appearance and strength of the crystalline silicate
features, the use of the ``Suh-silicate'' will not affect our results. 
As the peak position of the crystalline
features in laboratory samples
is strongly dependent on the Fe-content, ISO observations
of AGB and post-AGB stars
demonstrate that the dust shells only contain Mg-rich crystalline 
silicates \citep{MWT_99_afgl4106,SKB_99_ohir}.
For these crystalline silicates, we used optical constants of two
different types: forsterite (Mg$_2$SiO$_4$) and enstatite (MgSiO$_3$)
\citep{JMD_98_crystalline}.  
It is important to realize that because of the low Fe-content, 
their absorptivity
in the Near Infrared (NIR) is very low.
Equal mass fractions of both crystalline
species have been used, unless stated otherwise.

\section{Model results}
\label{sec:results}

Here we present the model results for increasing 
mass-loss rates (Sect.~\ref{sec:mdot}) and increasing 
degree of crystallinity (Sect.~\ref{sec:xtal}). 
In Sect.~\ref{sec:limits} the observational constraints
of ISO will be taken into account and applied to the model spectra.

\subsection{Varying the mass-loss rate}
\label{sec:mdot}

\begin{figure*}
  \resizebox{12cm}{!}{\includegraphics{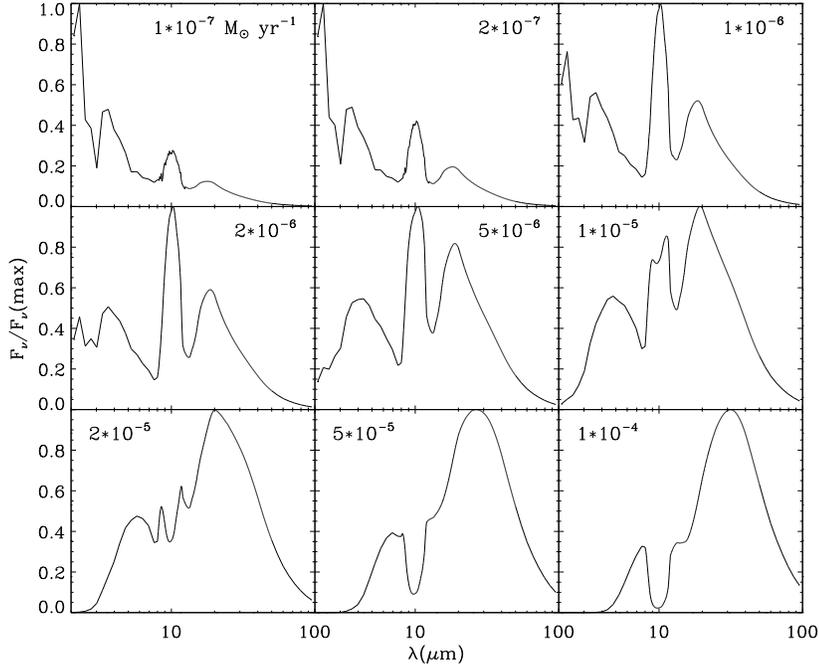}}
  \hfill \parbox[b]{55mm}{
    \caption{Fluxes 
      calculated for a completely amorphous dust shell, i.e.~ $x = 0$,
      for a wide range of mass-loss rates,  calculated for 
      $v_{\mathrm{exp}} = 20$ km s$^{-1}$ and $f = 0.01$. 
      In each panel, the
      mass-loss rate $\dot{M}$ ($M_{\odot}$ yr$^{-1}$) is indicated.
      The spectra are normalised on the maximum value in the
      2 -- 100 $\mu$m wavelength region.  Around 10 and 20
      $\mu$m broad features due to amorphous silicates are present in
      the spectra, occurring either in emission or absorption,
      depending on $\dot{M}$ (see
      text).  Note that in case of optically thin dust shells (low
      $\dot{M}$) the stellar spectrum with its characteristic molecular
      absorption bands is still visible at $\lambda <
      8$ $\mu$m, whereas for the optically thick case (high
      $\dot{M}$), the stellar flux at NIR wavelengths is completely
      absorbed by the dust.  }
    \label{kemperplot0}}
\end{figure*}

\begin{figure*}
  \resizebox{12cm}{!}{\includegraphics{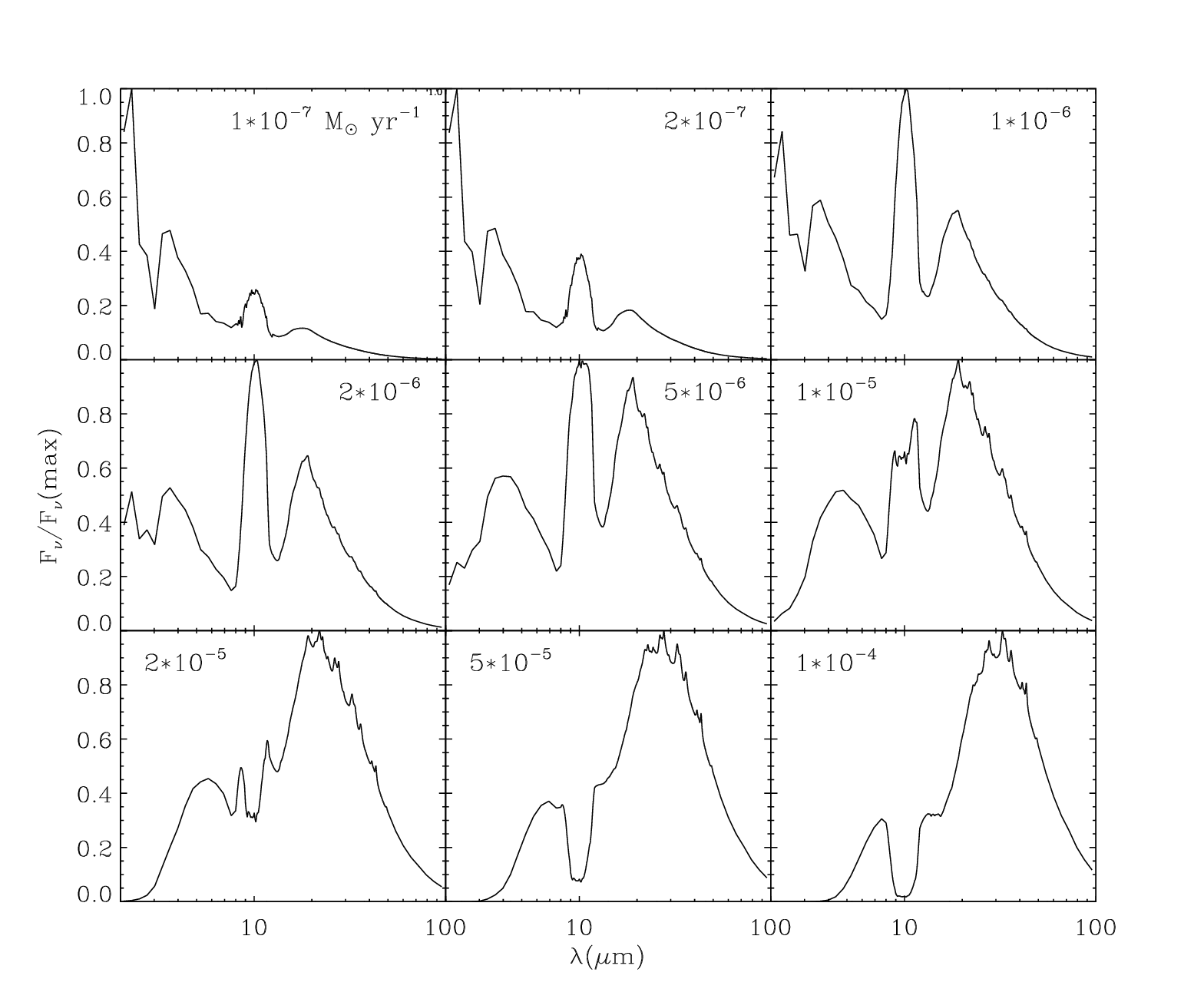}}
  \hfill \parbox[b]{55mm}{
    \caption{Normalised fluxes  
      calculated for a dust shell containing silicates with a degree
      of crystallinity $x = 0.10$, for the same mass-loss rates as
      in Fig.~\ref{kemperplot0}.  
      The narrow
      features due to crystalline silicates start to appear in the
      20 -- 50 $\mu$m range are only appearing for relatively high 
      mass-loss rates.  At $\dot{M} \approx 10^{-5}$
      $M_{\odot}$ yr$^{-1}$ the crystalline silicates are also
      discernible as substructure in the 10 $\mu$m feature (see also
      Fig.~\ref{closeup10})}
      \label{kemperplot10}}
\end{figure*}

\begin{figure*}
  \resizebox{12cm}{!}{\includegraphics{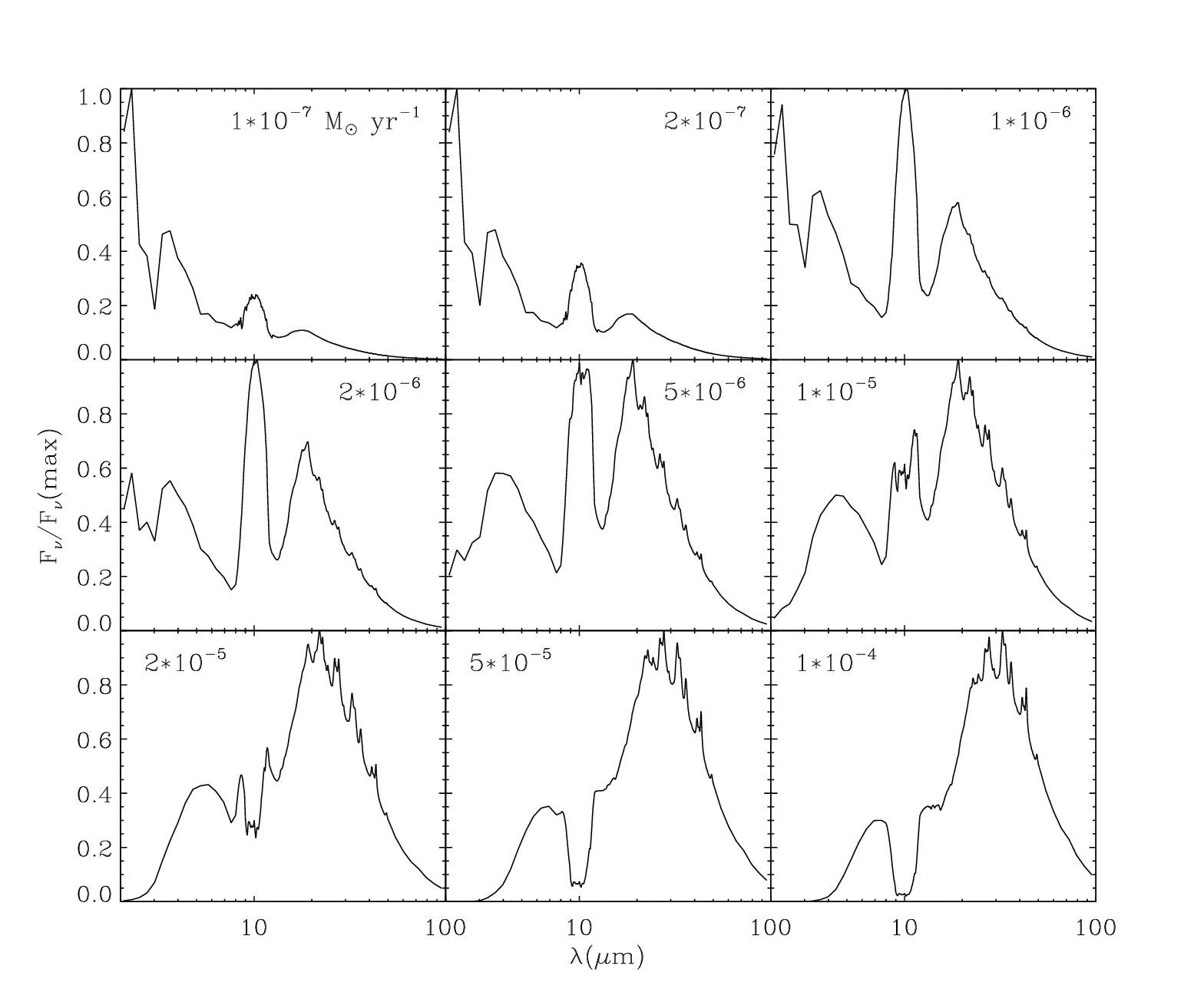}}
  \hfill \parbox[b]{55mm}{
    \caption{Normalised fluxes
      calculated for a dust shell containing silicates with a degree
      of crystallinity $x = 0.20$, for the same mass-loss rates as
      in Fig.~\ref{kemperplot0}.  The narrow
      features due to crystalline silicates in the 20 -- 50 $\mu$m
      range are only appearing for relatively high mass-loss rates.
      At $\dot{M} \approx 5 \cdot 10^{-6}$ $M_{\odot}$ yr$^{-1}$ the
      crystalline silicates are also discernible as substructure in
      the 10 $\mu$m feature (see also Fig.~\ref{closeup10})}
    \label{kemperplot20}}
\end{figure*}

In Figs.~\ref{kemperplot0}, \ref{kemperplot10} and \ref{kemperplot20}
results of the model calculations are presented. Each figure
shows the fluxes $F_{\nu} (\lambda[\mu \textrm{m}])$ for mass-loss rates
varying from 10$^{-7}$ to 10$^{-4}$ $M_{\odot}$ yr$^{-1}$ for a
constant degree of crystallinity. In Fig.~\ref{kemperplot0} 
results for a dust shell consisting of completely amorphous dust are
shown.  Clearly visible are the bands due to Si-O stretching and
O-Si-O-bending around 10 and 20 $\mu$m respectively. The strength of these
bands varies with increasing mass-loss rate; for low $\dot{M}$ the
dust shell is optically thin and the amorphous silicate bands appear
in emission. The strength of the emission bands increases with
increasing mass-loss rate. When the radial optical depth of the peaks of the
spectral
features approaches unity, the features become self-absorbed.
This occurs around $\dot{M} = 10^{-5}$ $M_{\odot}$ yr$^{-1}$ in case
of the 10 $\mu$m feature.  For the 20 $\mu$m feature this transition
is less well defined, and occurs at a somewhat higher mass-loss rate.
  
Figs.~\ref{kemperplot10} and \ref{kemperplot20} show the emerging
spectra for a crystallinity of 10\% and 20\% respectively. The overall
shape of these spectra resemble that of the amorphous silicate dust
shell spectra in Fig.~\ref{kemperplot0}. However, while the degree
of crystallinity remains constant,  
the narrow crystalline silicate features at 23.6, 27.6, 32.5,
36.1, 41.2 and 43.2 $\mu$m become quite apparent at relatively high mass-loss
rates.

\begin{figure}
  \resizebox{\hsize}{!}{\includegraphics{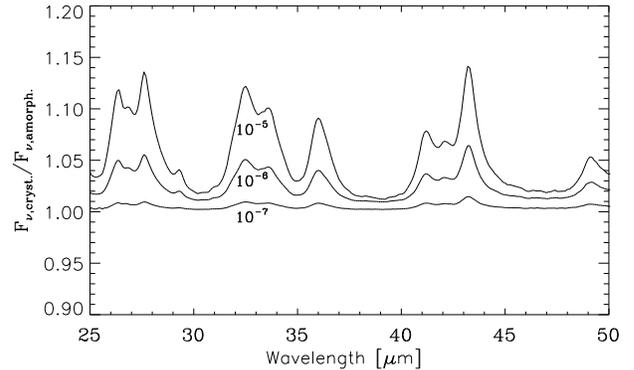}}
  \caption{Contrast plot of the 25 -- 50 $\mu$m region 
    for different mass-loss rates at a constant degree of
    crystallinity $x = 0.10$. The $\dot{M}$
    ($M_{\odot}$ yr$^{-1}$) values are indicated for each curve. The contrast
    is obtained by dividing the model spectrum containing 10\% crystalline
    dust by the model spectrum of a dust shell with an equal amount of 
    amorphous dust, but lacking the crystalline
    dust component.  }
  \label{contrastx10}
\end{figure}

In the models with low mass-loss rate, the narrow crystalline
silicate features are not discernible. Let us first investigate
the relative strength of the crystalline silicate features with
respect to the amorphous dust continuum in some more detail.  For
this purpose the contrast is calculated by dividing the spectrum of a
dust  model of a given crystallinity by the spectrum of a model
with equal properties and the same amorphous silicate dust mass, but
which is lacking the mass component taken up by the crystalline
silicates. Fig.~\ref{contrastx10} shows the contrast for three
different mass-loss rates, including 10\% crystalline material.  
The narrow bands due to crystalline silicates at 27.6, 32.5,
36.1, 41.2 and 43.2 $\mu$m show that the relative strength increases with
mass-loss rate. Note that the continuum between the
crystalline silicate features does not necessarily return to
unity, as the ''extra'' IR emission from crystalline silicates
raises the temperature of the amorphous silicates slightly. 
For $\dot{M} = 10^{-7}$
$M_{\odot}$ yr$^{-1}$ the contrast between the crystalline silicate
peaks and the surrounding continuum is less than 1\% for all features;
for $\dot{M} = 10^{-6}$ $M_{\odot}$ yr$^{-1}$ some 
peaks have a relative strength of $\sim 4$\% above continuum;
whereas the strongest peaks of the $\dot{M} = 10^{-5}$ $M_{\odot}$
yr$^{-1}$ model show a contrast of $\sim 10$\% above continuum.
The detection limit of ISO for broad spectral features in this
wavelength region is --- depending on the quality of the spectrum ---
around 5\% of the continuum level, which implies that of the three
curves plotted in Fig.~\ref{contrastx10} only the ISO spectrum of a
star with an outflow of 10$^{-5}$ $M_{\odot}$ yr$^{-1}$ would result
in a unambiguous detection of crystalline silicate features. A
detection in the spectrum of an AGB star with $\dot{M} = 10^{-6}$
$M_{\odot}$ yr$^{-1}$ would be dubious. A typical Mira, with
$\dot{M} = 10^{-7}$ $M_{\odot}$ yr$^{-1}$, with a crystallinity of
10\% would not show the characteristic features. This result is
consistent with the threshold mass-loss rate above which crystalline
silicates are detected with ISO observations
\citep{WMJ_96_mineralogy,CJJ_98_ohir,SKB_99_ohir}.

\subsection{Varying the degree of crystallinity}
\label{sec:xtal}
 
\begin{figure}
  \resizebox{\hsize}{!}{\includegraphics{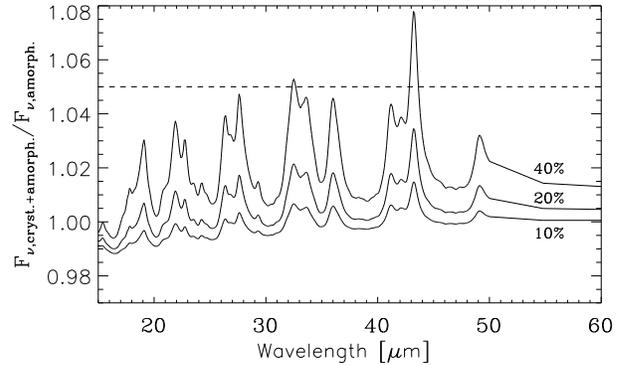}}
  \caption{Contrast plot for $\lambda < 60$ $\mu$m 
    for different degrees of crystallinity $x$ at a constant mass-loss
    rate of $\dot{M} = 10^{-7}$ $M_{\odot}$ yr$^{-1}$. The
    crystalline fraction $x$ is indicated at each curve. The contrast is
    obtained by dividing the model spectrum of a dust shell with a
    certain degree of crystallinity by the model spectrum of a dust
    shell with exactly the same amount of amorphous dust, but which
    lacks the crystalline dust component. The ISO detection limit for
    emission features of 5\% with respect to a continuum at $F_{\nu ,
      \mathrm{cryst+amorph}}/F_{\nu , \mathrm{amorph}} = 1$ is indicated with
    a dashed line.}
  \label{contrastm1e7}
\end{figure}

In order to determine the amount of crystalline silicates that should
be present in a typical Mira dust shell such that the narrow features
would be detected in an ISO spectrum, model spectra with different degrees of
crystallinity have been calculated. For a typical Mira mass-loss
of $\dot{M} = 10^{-7}$ $M_{\odot}$ yr$^{-1}$ the crystallinity has
been varied from 5\% to 50\%. The 
spectra of the $x=0.10$ and $x=0.20$
models, discussed in Sect.~\ref{sec:mdot}, are included in the overview.  
The contrast has been calculated in the
same way as described above. The result is
shown in Fig.~\ref{contrastm1e7}, for models with $\dot{M} = 10^{-7}$
$M_{\odot}$ yr$^{-1}$.  The dashed line indicates the 5\% detection
limit of ISO for solid state features, typical for $\lambda \sim$ 25 -- 50 
$\mu$m, and thus gives an impression of
which features will be discernible. As stated before, the relative
flux of the crystalline silicate model in between the narrow
crystalline silicate features is not necessarily equal to that of the
amorphous model fluxes.
The peak strength should therefore be compared to the local continuum,
rather than to unity. For this purpose an additional 
base-line has
to be subtracted. Following this method one easily sees that very high mass
abundances of crystalline silicates are required for a marginal
detection of crystalline silicate features in a typical Mira spectrum.
For a crystallinity of $x = 0.40$ only the 43.2 $\mu$m  and 32.5
$\mu$m features can be
detected by ISO.

\subsection{A trend in crystallinity?}
\label{sec:limits}

From Sects.~\ref{sec:mdot} and~\ref{sec:xtal} it becomes clear that
the comparison of the ISO detection limit with the contrast of the
crystalline silicate features is essential for studying the
correlation between mass-loss rate and crystallinity of AGB stars.
From Fig.~\ref{contrastm1e7} the 43.2 $\mu$m feature due to
crystalline pyroxene (enstatite, MgSiO$_3$) 
has been selected for closer examination, which
is a conservative choice.  It shows the largest contrast with the
local continuum, and would therefore be the first feature to be
detected in the spectrum when mass loss or crystallinity increases,
under the assumption that the mass fractions of crystalline pyroxene
and forsterite are equal.

\begin{figure}
  \resizebox{\hsize}{!}{\includegraphics[angle=0]{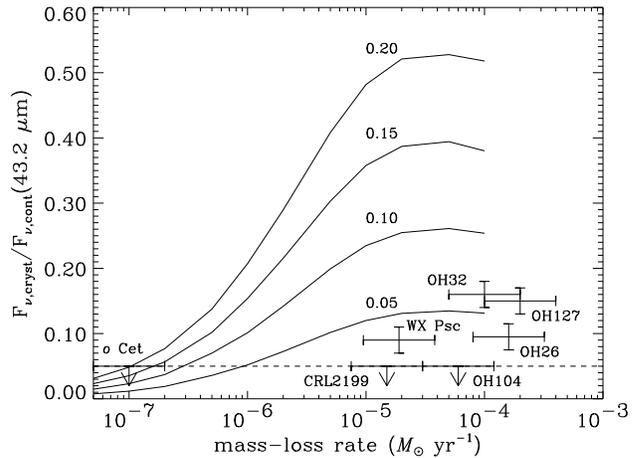}}
  \caption{Relative strength of the 43.2 $\mu$m feature with respect to the
local continuum as determined by amorphous material 
as a function of $\dot{M}$. The curves are calculated for
enstatite mass fractions of 0.05; 0.10; 0.15 and 0.20; which are indicated
in the plot. The dashed line represents the ISO detection limit at 43.2 
$\mu$m, which is $\sim$ 5\%. For several AGB stars, $\dot{M}$ and 
$F_{\nu,\mathrm{cryst}}/F_{\nu,\mathrm{cont}}$ are indicated (see 
Table \ref{stars}). For some stars,
only an upper limit to the relative flux is given.
 }
  \label{reversed43}
\end{figure}

For a range of enstatite mass fractions and mass-loss rates the
contrast spectra have been calculated. To determine the relative peak
strength with respect to the local continuum, a first order polynomial
base-line has been subtracted from the feature, such that the local
continuum is equal to unity. Then the peak value of the 43.2 $\mu$m
feature was measured, and plotted in Fig.~\ref{reversed43}. Note that the
enstatite mass fractions differ a factor 2 with
the elsewhere used crystallinity $x$, because it was assumed in
Sect.~\ref{sec:optconst} that enstatite and forsterite are equally
abundant. Fig.~\ref{reversed43} is a diagnostic tool to determine
the crystalline pyroxene content of the silicates present in the
dust shell around an AGB star. To do so, one has to determine
the relative peak strength of the 43.2 $\mu$m feature with respect
to the local continuum, a quantity plotted on the vertical axis,
and combine this with the mass-loss rate of the object. 
The data
points of the sample of AGB stars previously analysed by \citet{SKB_99_ohir} 
are presented to illustrate the method. The 
relative strength is measured from 
ISO spectra with an accuracy of 2 per cent of the continuum
(Table~\ref{stars}). We adopted an
accuracy of a factor 2 for the mass-loss rates.  
Using these data, one can estimate from Fig.~\ref{reversed43}
that the enstatite makes up around 3\% $\pm$ 2\% of the total silicate
dust mass around \object{WX~Psc} and \object{OH~26.5+0.6}, and around 
6\% $\pm$ 2\% in
\object{OH~127.8+0.0} and \object{OH~32.8$-$0.3}. In \object{CRL~2199} and 
\object{OH~104.9+2.4} the
enstatite mass fraction is well below 2--3\%, 
whereas the upper limit obtained for \object{$o$~Cet} (Mira itself)
does not put interesting constraints on the enstatite content. 
The dust shell of \object{$o$~Cet} could still easily contain 
a mass fraction of 20\% of crystalline pyroxene.
A constant enstatite mass fraction 
of $\sim$ 5\% is consistent with the observed strength
of the 43.2 $\mu$m feature for a wide range of mass-loss rates.

\begin{table}
\caption{Relative fluxes and mass-loss rates of the sample stars
from \citet{SKB_99_ohir}}
\label{stars}
\begin{tabular}{lcccc}
\hline
\hline
Star          & rel.~flux    & rel.~flux    & $\dot{M}$             & ref.\\

              &  32.5 $\mu$m  & 43.2 $\mu$m   & $M_{\odot}$ yr$^{-1}$ & \\
\hline
\object{$o$~Cet}       & $< 0.05$         & $< 0.05$         & $1.0 \cdot 10^{-7}$   & 1\\
\object{CRL~2199}      & $< 0.05$         & $< 0.05$         & $1.5 \cdot 10^{-5}$   & 2,3\\
\object{WX~Psc}        & 0.115  & 0.090  & $1.9 \cdot 10^{-5}$   & 2,4,5,6\\
\object{OH~104.9+2.4}  & 0.064  & $< 0.05$         & $6.0 \cdot 10^{-5}$   & 6\\
\object{OH~127.8+0.0}  & $< 0.05$         & 0.15  & $2.0 \cdot 10^{-4}$   & 3\\
\object{OH~26.5+0.6}   & 0.062  & 0.095 & $1.6 \cdot 10^{-4}$   & 3,6,7\\ 
\object{AFGL~5379}     & 0.077  & 0.093 &                       & \\
\object{OH~32.8$-$0.3} & $< 0.05$         & 0.16 & $1.0 \cdot 10^{-4}$   & 3,6,8\\   
\hline
\hline
\end{tabular}
\\
1~\cite{PBM_90_mira} 2~\cite{KM_85_massloss} 3~\cite{JT_92_SED} 4~\cite{SOJ_89_molemm} 5~\cite{LW_98_IRMdot} 6~\cite{ST_89_theory} 7~\cite{JST_96_OH26} 8~\cite{G_94_OH32OH44}
\end{table}

\begin{figure}
  \resizebox{\hsize}{!}{\includegraphics[angle=0]{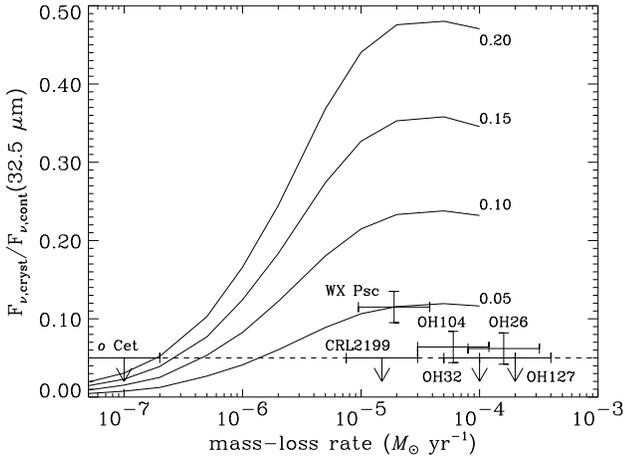}}
  \caption{Relative strength of the 32.5 $\mu$m forsterite feature with 
respect to the local continuum as a function of $\dot{M}$. 
The curves are calculated for forsterite mass fractions of 
0.05; 0.10; 0.15 and 0.20; which are indicated
in the plot. The dashed line represents the ISO detection limit at 32.5 
$\mu$m, which is $\sim$ 5\%. For a sample of AGB stars, $\dot{M}$ and 
$F_{\nu,\mathrm{cryst}}/F_{\nu,\mathrm{cont}}$ are indicated (see
Table~\ref{stars}). For some stars, only an
upper limit to the relative strength is available. }
  \label{reversed33}
\end{figure}

A similar analysis can be performed on the forsterite
(Mg$_2$SiO$_4$, crystalline olivine) mass fraction, 
using the 32.5 $\mu$m feature,
which has the best contrast. For the same range of mass-loss rates 
the emerging dust spectrum has been calculated. 
The mass fraction of the crystalline
olivine varied from
0\% to 20\%. All other input parameters had equal values as 
the parameters in the
analysis of the crystalline pyroxene mass fraction.
We calculated the relative peak value of the 32.5 $\mu$m feature,
by subtracting a first order polynomial from the forsterite contrast
plots.
Fig.~\ref{reversed33} shows the relative peak fluxes as a function of
mass-loss rate, calculated for different mass fractions of forsterite.
From this plot it is obvious that in low mass-loss rate AGB stars,
the strength of the 32.5 $\mu$m feature is less than 5\% of the continuum,
which makes it imperceptible for ISO, as is illustrated with the data
for $o$ Cet.
For higher mass-loss rate AGB stars,
the strength of the feature provides a reliable measure to determine
the crystalline olivine mass fraction. The forsterite mass fraction
of our
sample of high mass-loss stars however, is at most 5\% $\pm$ 2\%
for WX~Psc. There is no evidence for an increasing forsterite mass
fraction for higher mass-loss rates.

\begin{figure}
  \resizebox{\hsize}{!}{\includegraphics[angle=0]{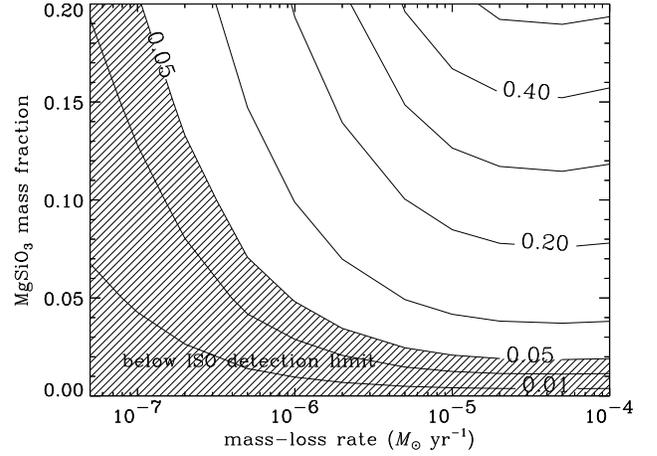}}
  \caption{Relative peak strength of the 43.2 $\mu$m feature as 
    a function of mass-loss rate and fraction of mass taken by
    enstatite (MgSiO$_3$). The peak strength is
    determined with respect to the local continuum.  A relative peak strength
     $> 5$\% is observable with ISO. The shaded
    area represents the values for $\dot{M}$ and enstatite mass
    fraction for which the strength of the 43 $\mu$m feature is below
    the ISO detection limit. The mass fraction of enstatite is sampled
    at 0.00, 0.05, 0.10 and 0.20; and the mass-loss rate is sampled
    at $\dot{M}/10^6$ =
    0.05, 0.1, 0.2, 0.5, 1, 2, 5, 10, 20, 50 and  100 $M_{\odot}$ yr$^{-1}$. }
  \label{contour43}
\end{figure}

\begin{figure}
  \resizebox{\hsize}{!}{\includegraphics[angle=0]{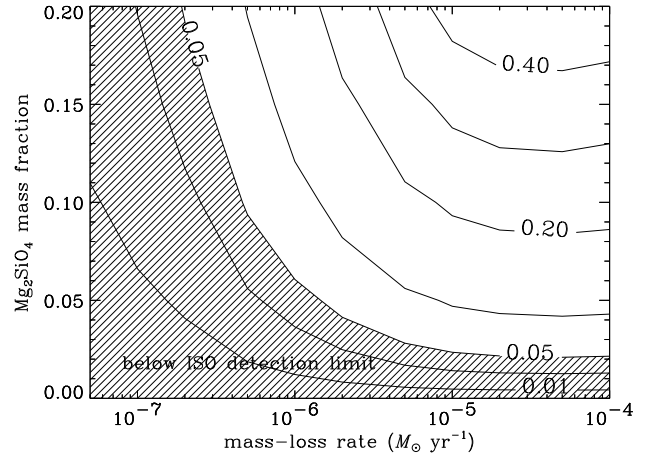}}
  \caption{Relative peak strength of the 32.5 $\mu$m feature as 
    a function of mass-loss rate and fraction of mass taken by
    forsterite (Mg$_2$SiO$_4$, crystalline olivine). The peak strength is
    determined with respect to the local continuum.  A relative peak strength
     $> 5$\% is observable with ISO. The shaded
    area represents the values for $\dot{M}$ and forsterite mass
    fraction, for which the strength of the 32.5 $\mu$m feature is below
    the ISO detection limit. The mass fraction of forsterite is sampled
    at 0.00, 0.05, 0.10, 0.15 and 0.20; and the mass-loss rate is sampled
    at  $\dot{M}/10^6$ =
    0.05, 0.1, 0.2, 0.5, 1, 2, 5, 10, 20, 50 and  100 $M_{\odot}$ yr$^{-1}$. }
  \label{contour33}
\end{figure}

Figs.~\ref{contour43} and \ref{contour33} illustrate 
that an observational selection
effect is present in the correlation between mass-loss rate and
crystallinity reported in previous studies by 
\citet{WMJ_96_mineralogy,CJJ_98_ohir,SKB_99_ohir}. 
The OH/IR stars with $\dot{M} \gtrsim 10^{-5}$ $M_{\odot}$
yr$^{-1}$ only have to contain a small mass fraction of crystalline
pyroxene in their dust shells in order to exhibit the 43.2 
$\mu$m or 32.5 $\mu$m feature. A few percent is already perceptible with 
ISO. However, the
43.2 and 32.5 $\mu$m 
features in the spectra of Miras, with $\dot{M} \lesssim 10^{-6}$
$M_{\odot}$ yr$^{-1}$, will only be detected for 
high crystallinity. 
For example, in the dust shell of $o$ Cet, 
with $\dot{M} \approx 10^{-7}$ $M_{\odot}$ yr$^{-1}$, a
dust mass fraction of 20\% of enstatite could be present without
showing the 43 $\mu$m feature in its spectrum. The same goes for
the 32.5 $\mu$m forsterite feature. If we assume that equal
amounts of crystalline pyroxene and olivine are present, the ISO
detection limit only provides an upper-limit to the crystallinity of
$x = 0.40$.  If a constant crystallinity of $x=0.10$ (i.e.~a mass
fraction for both enstatite 
and forsterite of 5\% of the total silicate mass) is assumed,
while the star evolves along the AGB, the 32.5 $\mu$m and 
43.2 $\mu$m  features will
appear in the spectrum when the mass loss is sufficiently high. The threshold
value, defined by the ISO detection limit, is found at $\dot{M} \sim
10^{-6}$ $M_{\odot}$ yr$^{-1}$.  This is consistent with 
observational studies that conclude that crystalline silicate features
only appear to be present in high mass-loss rate objects.

\section{Physical explanation and discussion}
\label{sec:discussion}

The appearance of crystalline silicate features with increasing
mass-loss rate for a constant degree of crystallinity may be caused 
by the difference in absorptivity in the NIR of crystalline and
amorphous silicates.  This difference is due to a difference in
Fe-content of the amorphous and crystalline components. The
crystalline silicates investigated do not contain any iron, whereas in the
amorphous silicates the iron and magnesium contents are roughly equal.
Previous laboratory 
studies \citep{KST_93_olivpyro,JMD_98_crystalline} have shown
that the peak positions of crystalline silicate features strongly
depend on the Fe-content of the silicate. At present, there have not
been any detections of crystalline silicates containing iron around
evolved stars, which justifies the use of laboratory data of pure
Mg$_2$SiO$_4$ and MgSiO$_3$.

The absorptivity in the NIR and at shorter wavelengths 
increases with increasing
Fe-content. 
The amorphous silicates thus absorb more radiation from the central star,
which emits most of its flux at NIR wavelengths, 
than the crystalline
silicates do, resulting in a large temperature difference between
amorphous and crystalline material, the latter staying much cooler.
In the mid and far infrared the absorptivity of both dust
types are more or less equal. In an optically thick dust shell, where
the stellar flux is absorbed and re-radiated several times by the
dust, the grains in the outer regions are illuminated by a
radiation field that typically 
peaks at mid infrared wavelengths. As in this wavelength regime only a small
difference in absorptivity occurs, a large temperature difference
between the amorphous and crystalline dust components will not
develop.  Summarising, at the inner radius of an optically thick dust shell the
crystalline silicates are much colder than the amorphous silicate
grains, but the temperature difference decreases at larger distances.

The implications for the strength of the crystalline silicate features
due to this temperature difference are quite dramatic.  In low
mass-loss rate AGB stars (Miras), the dust shell is optically thin, 
implying the entire dust shell is visible and contributes to the
SED. The relatively hot amorphous silicates at the inner radius then
dominate the SED of a Mira, and the contrast of the crystalline
silicates with respect to the amorphous silicate spectrum becomes very
poor.  In case of a high mass-loss rate AGB star, only the outer
regions of the dust shell are visible, because the shell is optically
thick at IR wavelengths. The temperature differences in the outer
layers are small, as explained above, 
and therefore the crystalline and amorphous dust
components both contribute in comparable amounts to the spectrum. This
improved contrast of the crystalline silicates with respect to the
amorphous dust emission enables the detection of the narrow
crystalline features in the spectrum of OH/IR stars.

The ISO detection limit allows the presence of a significant
fraction of crystalline silicates in the circumstellar shells of low
mass-loss rate AGB stars. In fact, the ISO data are consistent with
model calculations assuming that the crystallinity of the dust is 
constant with mass-loss.  Until now, most theoretical  studies
\citep{TWM_98_mineralogy,GS_99_condensation,SK_99_xsilagb} 
assumed that crystalline
silicates are only formed in high density outflows.  The results
presented in this study suggest that the possibility of the
condensation of crystalline silicates in low mass-loss rate AGB stars
should be considered in future studies.

\subsection{How to search for crystalline silicates in Miras}
\label{sec:obsevidence}

\begin{figure}
  \resizebox{\hsize}{!}{\includegraphics[angle=90]{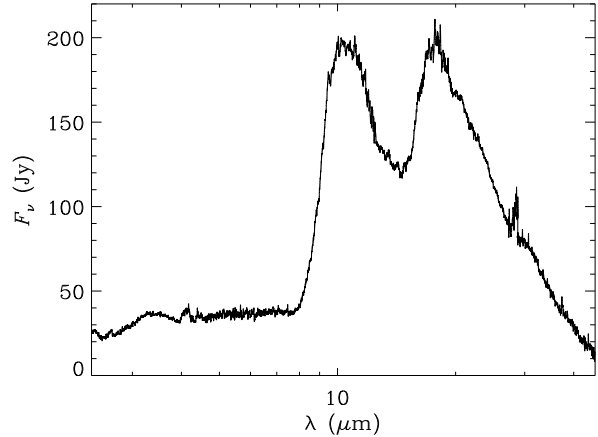}}
  \caption{ISO SWS spectrum of \object{AFGL~2999}. The spectrum of this star is
typical for that of an object somewhere in between typical Miras and
OH/IR stars. High resolution and high signal-to-noise observations may
perhaps reveal crystalline silicate features as sub-structure in the
broad amorphous 10 $\mu$m feature. No crystalline silicates are seen at 
the 5 percent level in the 20 -- 45 $\mu$m range. The artefact at 28
$\mu$m is due to calibration problems with band 3e of ISO SWS.}
  \label{afgl2999}
\end{figure}

To date, there are no significant detections of crystalline silicates
in low mass-loss rate AGB stars reported. As shown above, 
this provides a relatively
high upper limit to the mass fraction taken by crystalline silicates
in the shells around these stars. In order to lower this upper limit
and to investigate the presence of crystalline silicates in Miras,
additional observations with a better detection limit than that of
ISO, are required. Unfortunately, the most prominent and
characteristic features of crystalline silicates appear in the
25 -- 50 $\mu$m region, and with the end of the ISO mission the
opportunities to observe in this wavelength region are limited. 
An important future
mission covering part of this wavelength region is SIRTF, which can
perform intermediate resolution spectroscopy ($R \sim 600$) up to $\lambda
\approx 37$ $\mu$m. The crystalline silicate feature with the largest
contrast value within this wavelength range is the 32.5 $\mu$m
forsterite feature.  In Fig.~\ref{contour33}, the contrast of 
this feature with respect to the continuum is presented as a
function of mass-loss rate and Mg$_2$SiO$_4$ mass fraction.

\begin{figure}
  \resizebox{\hsize}{!}{\includegraphics{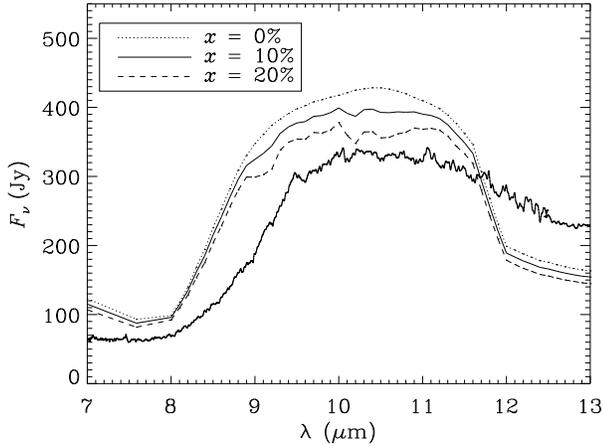}}
  \caption{Predicted flux levels in the 7 -- 13 $\mu$m 
    region for an AGB star with $\dot{M} = 5 \cdot 10^{-6}$ $M_{\odot}$
    yr$^{-1}$. Model spectra with three different crystallinities
    ($x=0.0$; $x=0.10$; $x=0.20$) are indicated. The broad emission
    feature is due to amorphous silicates. 
    The substructure is due to crystalline
    silicate features that appear in absorption. The ISO SWS spectrum of 
    \object{AFGL~2999} is plotted for reference; in this spectrum it is not
    possible to detect crystalline silicate features, since the 
    the substructure in the spectrum is mostly due to observational 
    noise.}
  \label{closeup10}
\end{figure}

In addition, it is also possible to study the 10 and 20 $\mu$m silicate
features, which both can be observed from the ground.  Since it is
already known that the high mass-loss rate OH/IR stars contain
crystalline silicates in their circumstellar dust shells, it is
particularly interesting to study the 10 $\mu$m feature of low and
intermediate mass-loss rate objects, with $\dot{M} < 10^{-5}$
$M_{\odot}$ yr$^{-1}$. From figs.~\ref{kemperplot10} and
\ref{kemperplot20} it one can conclude that the 10 $\mu$m feature of
intermediate mass-loss rate AGB stars ($\dot{M} \approx 10^{-6}$ to
$10^{-5}$ $M_{\odot}$ yr$^{-1}$) would be the most likely to reveal the
presence of crystalline silicates. The 10 $\mu$m region has the
intrinsically strongest resonance peaks of crystalline silicates. Superposed
on the 10 $\mu$m feature of amorphous silicates which is just optically
thick ($\tau \gtrsim 1$) these strong resonances appear as absorption features
in the spectrum.  The intermediate mass-loss rate
AGB stars can be easily recognised from the ratio between the strength
of the 10 and 20 $\mu$m features. For example \object{AFGL~2999}, of which the
ISO SWS spectrum is shown in Fig.~\ref{afgl2999}, can be identified as
a $\dot{M}= 5 \cdot 10^{-6}$ $M_{\odot}$ yr$^{-1}$ AGB star.

Fig.~\ref{closeup10} shows predicted 10 $\mu$m features. The
flux levels are calculated for models with $\dot{M} = 5 \cdot 10^{-6}$
$M_{\odot}$ yr$^{-1}$ and $x = 0.00, 0.10, 0.20$.  The broad amorphous
silicate feature is partially optically thick and shows
a flattened profile due to
self-absorption. The additional opacity provided by the crystalline
silicates appears as narrow absorption superposed on the broad
amorphous silicate feature. The ISO spectrum of \object{AFGL~2999} is plotted
for reference, but the substructure in this spectrum is observational
noise and it is therefore hard to disentangle the crystalline silicate
mass fraction. However, 
it would certainly be possible reach a high $S/N$ ratio 
and detect a
crystallinity of 10\% with the current ground-based spectrometers that
cover the 10 $\mu$m region, such as TIMMI2 mounted on the 3.6m ESO 
telescope.

\subsection{Implications for the dust formation and processing in the ISM}
\label{sec:dustform}

AGB stars are considered
the main contributors of dust to the ISM, so one would
expect a match between the dust composition in the ISM and in the
shells of AGB stars. However, despite all evidence for crystalline
silicates in AGB stars, there has as yet not been found any reliable
evidence for
crystalline silicates in the line of sight toward the Galactic Center
(GC) \citep{LFG_96_GC}, or other lines of sight through the ISM
\citep{DJD_99_dustcomposition}. In fact, the line of sight toward the
GC is found to resemble closely the silicate feature of the evolved
star $\mu$ Cep,
and can be fitted with completely amorphous olivine
\citep[][Bouwman et al.~\emph{in prep}]{V_99_10micronGC}. 
\citet{CJL_00_xsil_in_orion} report a questionable detection of 
crystalline silicates in the Orion Bar.
 
On the other hand, it is doubtful that Miras contribute significantly
to the dust population in the ISM. It is generally accepted that the
Mira phase lasts $\sim 3 \cdot 10^5$ yr, and the OH/IR phase $\sim 3000$ yr
\citep{H_96_review}. The mass-loss rates for Miras and OH/IR stars
($\dot{M}_{\mathrm{Mira}} \approx 10^{-7}$ $M_{\odot}$ yr$^{-1}$, and
$\dot{M}_{\mathrm{OH/IR}} \approx 10^{-4}$ $M_{\odot}$ yr$^{-1}$) can
be used to estimate the total mass lost in each of the two phases; 
for OH/IR stars this is $\sim
0.3 M_{\odot}$, for Miras $\sim 0.03 M_{\odot}$.  Under the 
assumption that all Miras will eventually evolve into OH/IR stars,
only a minor fraction ($\sim 10$\%) of the interstellar dust
population is formed in a Mira dust shell.  So, the possibility that Miras
contain a significant fraction of crystalline silicates in their dust
shells does not heavily aggravate the discrepancy between the dust
composition in the ISM and that in AGB dust shells, under the 
assumption that all Miras eventually evolve into OH/IR stars. 
However, if Mira stars indeed deposit dust with a high degree of crystallinity
into the ISM, and a significant fraction of Miras does not evolve into
OH/IR stars, it is more surprising that the interstellar
silicate grains appear to have an amorphous lattice structure.
Grain-grain collisions and ion 
bombardments might cause the required amorphitization. Due to 
the collision, the grain will partially melt and consequently
solidify in an amorphous form 
on timescales shorter than the annealing timescales
\citep{SKB_99_ohir}. In a
future study (Kemper et al., \emph{in prep.}) we will thoroughly 
examine the differences between silicates in
the ISM and in circumstellar shells, and discuss possible mechanisms
of grain processing.

\section{Summary}
\label{sec:summ}

In this work, three important aspects of crystalline silicates
around AGB stars have been studied by modelling the dust shell.
First, we have calculated spectra emerging from circumstellar
dust shells for different mass-loss rates. Moreover, the effects of different
degrees of crystallinity on the SED have been taken into account,
as a function of optical depth toward the central star. Finally, we provide
a useful and easy to use diagnostic tool to determine the degree of 
crystallinity for an AGB star with a known mass-loss rate.
 
The dust shells of low mass-loss rate AGB stars can contain a
significant fraction of crystalline silicates, while the
characteristic sharp spectral peaks are indiscernible with ISO. This
is caused by a large temperature difference in the inner parts of
these optically thin dust shells, such that the relatively warm
amorphous silicate
emission dominates the SED. The temperature difference between the
warm amorphous dust and the cold crystalline dust is caused by a
difference in absorptivity at NIR and visible wavelengths, where
the amorphous dust component more efficiently absorbs the stellar
radiation. The low absorptivity of the crystalline component is due to the
very low Fe-content of these dust species. 
In case of high mass-loss AGB stars, the
radiation from the central star is absorbed by the dust and re-radiated
several times, so that the dust in the outer parts receives a 
relatively red radiation field, which peaks in the mid-IR. In that region
the absorptivity of crystalline and amorphous silicates are similar,
leading to similar temperatures of both species. The dust shell
is optically thick, so only the outer layers, where crystalline and
amorphous silicates have similar temperatures, are visible, leading
to a high contrast of the crystalline silicate features.

From these results one may conclude that crystallinity is not
necessarily a function of mass-loss rate, as many observational and
theoretical studies have suggested. The threshold value for the
mass-loss rate above which crystalline silicates are observed 
is consistent with the ISO detection limit if a constant
crystallinity of 10\% is assumed. The strong observational selection
effect undermines previously drawn conclusions regarding this threshold value.
The meaning and the existence of the threshold mass-loss rate should
be reconsidered. Future observations of
the 10 $\mu$m silicate feature of low mass-loss AGB stars
at high spectral resolution  can
probably further constrain the upper limit for the crystalline
silicate content in Miras. The low degree of crystallinity of grains
in the interstellar medium with respect to grains in AGB shells, however,
remains subject to further study.

\begin{acknowledgements}
  We wish to thank C. Dominik, J. Bouwman, F.J. Molster and T. de Jong for the
  useful discussions. We are grateful to I. Yamamura for the reduction
  of the ISO SWS spectrum of \object{AFGL~2999}. FK, LBFMW and AdK
  acknowledge financial support from NWO Pionier grant 
  616-78-333. We gratefully acknowledge support from NWO Spinoza
  grant 08-0 to E.P.J.~van den Heuvel.
\end{acknowledgements}

\bibliographystyle{apj}
\bibliography{ciska}

\end{document}